\documentclass[10.5pt]{article}
\usepackage{amsfonts}
\usepackage{amsmath}
\usepackage{amssymb}
\usepackage{amstext}
\usepackage{amsfonts}
\usepackage{graphicx}
\usepackage{indentfirst}
\usepackage{hyperref}
\usepackage{cite}
\usepackage{epsfig}
\usepackage{CJK}
\makeatletter
\def\ExtendSymbol#1#2#3#4#5{\ext@arrow 0099{\arrowfill@#1#2#3}{#4}{#5}}
\def\RightExtendSymbol#1#2#3#4#5{\ext@arrow 0359{\arrowfill@#1#2#3}{#4}{#5}}
\def\LeftExtendSymbol#1#2#3#4#5{\ext@arrow 6095{\arrowfill@#1#2#3}{#4}{#5}}
\makeatother

\begin{document}

%=================== Text begin here =============================================

\begin{center}
\LARGE\bf Synchronous concentration and purification schemes of
arbitrary unknown hyperentangled mixed states   %% ÂÛÎÄÌâÄ¿
\end{center}

\footnotetext{\hspace*{-.45cm}\footnotesize $^\dag$Corresponding author, E-mail: qiaocf@ucas.ac.cn  }

\begin{center}
\rm Kun Du, Qiucheng Song, Congfeng Qiao$^{\dagger}$

\end{center}

\begin{center}
\begin{footnotesize} \sl
(University of Chinese Academy of Sciences, Beijing 100049,
China)
\end{footnotesize}
\end{center}
\begin{center}
%\footnotesize (Received X XX XXXX; revised manuscript received X XX XXXX)
          %% (Received ÈÕ ÔÂ Äê; revised manuscript received ÈÕ ÔÂ Äê)
\end{center}
\vspace{2mm}

\begin{center}
\begin{minipage}{15.5cm}
\begin{minipage}[t]{2.3cm}{\bf Abstract}\end{minipage}%%%% ÂÛÎÄÕªÒª
We present two efficient schemes which can simultaneously accomplish hyperentanglement concentration and purification for two-photon four-qubit systems in an
unknown partially hyperentangled mixed states. The first can correct errors in the polarization entanglement and extract maximal hyperentanglement in polarization and
spatial mode with additional partial frequency entanglement. The second uses additional maximal frequency entanglement to purify and concentrate hyperentanglement
in polarization and spatial mode deterministically. Both of the two schemes are only
based on existing optical devices and cross-Kerr nonlinearities.
\end{minipage}
\end{center}

\begin{center}
\begin{minipage}{15.5cm}
\begin{minipage}[t]{2.3cm}{\bf Keywords}\end{minipage}%%%%% ¹Ø¼ü´Ê
hyperentanglement; concentration; purification; cross-Kerr nonlinearities
\end{minipage}
\end{center}

\section{Introduction}
Entanglement is viewed as a kind of raw resource of quantum
information science, such as measurement-based quantum computing
\cite{one}, quantum teleportation \cite{teleportation}, quantum dense coding \cite{superdense},
and quantum cryptography \cite{Cryptography,QKD}. Meanwhile, to go further
in the manipulation of more entangled qubits, hyperentanglement, namely making
the quanta, e.g. photons, to be entangled simultaneously in multiple
degrees of freedom (DOFs) has received more and more attention for quantum information
process \cite{Kwiat,Ultrabright,Polarization-momentum}. For example, in recent years, it has been applied in quantum key distribution (QKD)
protocol \cite{xu}, Bell-state analysis \cite{Bell-state,Bell1,Bell2}, entanglement purification protocol (EPP)
\cite{EPP1,EPP2,EPP3} and quantum repeater protocol \cite{repeater}.

Although at present the preparation of hyperentanglement is high-efficient and high-quality, the entangled photon
pairs are usually locally produced, thus decoherence in the long-distance quantum communication channel
is unavoidable, which will significantly reduce the quality of photon pairs and decrease their entanglement.
Therefore the efficiency and fidelity of quantum communication protocols between distant locations will be greatly
decreased. The main methods to overcome decoherence in quantum communication process are entanglement purification
and entanglement concentration. The former is a method by which one can obtain a
smaller set of high-fidelity entangled pairs from a large number of less-entangled pairs in a mixed state.
The latter is used to distill maximally entangled pairs
out of a set of partially entangled pairs in a pure state. In 1996, Bennett {\it et al.} proposed the first
entanglement purification protocol (EPP) for two-photon systems in mixtures of the four Bell states, resorting to quantum controlled-not
(CNOT) gates and local unitary operations \cite{EPP1996}. In 2002, Simon and Pan presented an EPP with parametric
down-conversion (PDC) source and currently available linear optical elements \cite{EPP2002}. In 2010, a deterministic EPP
with hyperentangled state was proposed by Sheng {\it et al.} \cite{EPP2010}. In 2013, Sheng {\it et al.} presented an EPP
to reconstruct some maximally hybrid entangled states from nonmaximally mixed systems  \cite{EPP2013}.
 Meanwhile, many significant entanglement concentration protocols (ECPs)
have been presented. For example, in 2012, Deng proposed an ECP for photon systems with known parameters based on projection measurements \cite{ECP2012}.
In 2001, an ECP with unknown polarization entangled states was proposed by Zhao {\it et al.} \cite{ECP2001}. Recently,
Ren {\it et al.} put forward a hyperentanglement concentration
protocol (hyper-ECP) for the systems in partially hyperentangled state \cite{ECP2013,ECP20132,ECP2014}.

In this work, we investigate the methods of simultaneously correcting errors and distilling maximal hyperentanglement
in both the polarization and spatial mode DOFs with two-photon systems in the nonlocal partially hyperentangled mixed states.
First, we only correct the bit-flip error and phase-flip error of the polarization entanglement, and extract
both the maximally polarization and spatial mode entangled states at the cost of additional partial frequency entanglement.
Subsequently, we simultaneously correct errors of polarization and spatial mode entanglement and extract the maximal hyperentanglement
by dint of additional maximal frequency entanglement.

\section{Hyper-ECP with additional partial frequency entanglement}
Our mission is to transmit the maximally Bell hyperentangled state $|\varphi\rangle=\frac{1}{2}(|H\rangle|H\rangle+|V\rangle|V\rangle)
\otimes(|a_{1}\rangle|b_{1}\rangle+|a_{2}\rangle|b_{2}\rangle)$ to the parties Alice and Bob. Using the SPDC source presented by Deng
{\it et al.} \cite{WDM}, we can produce pairs of photons entangled in three DOFs:
\begin{eqnarray}
|\psi\rangle&=&\frac{1}{2\sqrt{2}}(|HH\rangle+|VV\rangle)\nonumber\\
&&\otimes(|a_{1}b_{1}\rangle+|a_{2}b_{2}\rangle)\nonumber\\
&&\otimes (|\omega_{1}\omega_{2}\rangle+|\omega_{2}\omega_{1}\rangle),\label{shi2}
\end{eqnarray}
where $H$ and $V$ denote horizontal and vertical polarization,
$\omega_{1}$ and $\omega_{2}$ signify different frequency and
$a_{1}$, $b_{1}$, $a_{2}$, $b_{2}$ label different spatial modes, and suppose $\omega_{2}>\omega_{1}$.
As the spatial mode and frequency are more stable than polarization \cite{stable,unstable,bit-flip,phase-flip}, we assume that there are no bit-flip errors and phase-flip errors in the spatial mode and frequency DOFs, and they only become
partially entangled states after transmission through noisy channels:
\begin{eqnarray}
|\psi_{s}\rangle&=&\gamma|a_{1}b_{1}\rangle+\delta|a_{2}b_{2}\rangle,\\
|\psi_{f}\rangle&=&\varepsilon|\omega_{1}\omega_{2}\rangle+\eta|\omega_{2}
\omega_{1}\rangle.\label{shi2}
\end{eqnarray}
But the polarization state changes into a mixed one:
\begin{eqnarray}
\rho_{p}&=&F_{1}|\psi_{p_{1}}\rangle\langle\psi_{p_{1}}|+F_{2}|\psi_{p_{2}}
\rangle\langle\psi_{p_{2}}|\nonumber\\
&&+F_{3}|\psi_{p_{3}}\rangle\langle\psi_{p_{3}}|+F_{4}|\psi_{p_{4}}\rangle
\langle\psi_{p_{4}}|,\label{shi2}
\end{eqnarray}
where $F_{1}$+$F_{2}$+$F_{3}$+$F_{4}$=1, and
\begin{eqnarray}
|\psi_{p_{1}}\rangle=\alpha|HH\rangle+\beta|VV\rangle,\nonumber\\
|\psi_{p_{2}}\rangle=\alpha|HV\rangle+\beta|VH\rangle,\nonumber\\
|\psi_{p_{3}}\rangle=\alpha|HH\rangle-\beta|VV\rangle,\nonumber\\
|\psi_{p_{4}}\rangle=\alpha|HV\rangle-\beta|VH\rangle\;.\label{shi2}
\end{eqnarray}

We consider two pairs of photons AB and CD in the above mixed state. The photons A and C are transmitted to Alice,
and the photons B and D belong to Bob. The six parameters
$\alpha$, $\beta$, $\gamma$, $\delta$, $\varepsilon$, and $\eta$ are unknown to Alice and Bob, and they satisfy the relation
$|\alpha|^2+|\beta|^2=|\gamma|^2+|\delta|^2=|\varepsilon|^2+|\eta|^2=1$.

As the two pairs are both initially in the mixed state $\rho=\rho_{p}\rho_{s}\rho_{f}$, so there are 16 kinds of cases.
We will discuss the following situation as an example, which has both bit-flip error and phase-flip error, that is the initial state of the four-photon
system can be written as:
\begin{eqnarray}
|\phi\rangle&=&|\psi_{p_{1}}\rangle_{AB}|\psi_{s}\rangle_{AB}|\psi_{f}\rangle_{AB}\nonumber\\
&&\otimes|\psi_{p_{4}}\rangle_{CD}|\psi_{s}\rangle_{CD}|\psi_{f}\rangle_{CD} \nonumber\\
&=&(\alpha^2|HHHV\rangle-\beta^2|VVVH\rangle\nonumber\\
&&-\alpha\beta|HHVH\rangle+ \alpha\beta|VVHV\rangle) \nonumber\\
&&\otimes(\gamma^2|a_{1}b_{1}c_{1}d_{1}\rangle+\beta^2|a_{2}b_{2}c_{2}d_{2}\rangle\nonumber\\
&&+\gamma\delta|a_{1}b_{1}c_{2}d_{2}\rangle+\gamma\delta|a_{2}b_{2}c_{1}d_{1}\rangle)  \nonumber\\
&&\otimes(\varepsilon^2|\omega_{1}\omega_{2}\omega_{1}\omega_{2}\rangle+\eta^2|
\omega_{2}\omega_{1}\omega_{2}\omega_{1}\rangle\nonumber\\
&&+\varepsilon\eta|\omega_{1}\omega_{2}\omega_{2}\omega_{1}\rangle+\varepsilon\eta|
\omega_{2}\omega_{1}\omega_{1}\omega_{2}\rangle).\label{shi2}
\end{eqnarray}

\begin{figure}[h]
\centering
\includegraphics[width=8cm]{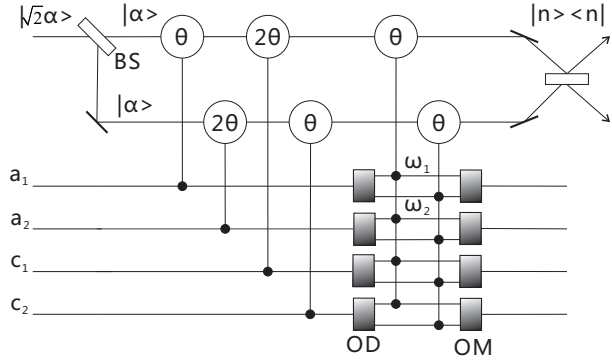}
\caption{$QND_{1}$. Extracting the parts of maximal entanglement in spatial mode and frequency with cross-Kerr nonlinear medium performed by Alice.
OD and OM denote an optical demultiplexer and an optical multiplexer respectively which are parts of wavelength division multiplexer (WDM).} \label{fig1}
\end{figure}

In the first place, Alice uses a quantum nondemolition detector (QND1) to pick out maximal entanglement in spatial mode and frequency.
As shown in Figure 1, photons A and C are led to a cross-Kerr nonlinear medium \cite{cross-Kerr,nonlinear}, which brings forth an adjustable phase shift to the coherent
states through cross-phase modulation (XPM). We use a 50:50 beam splitter (BS) to divide the coherent
state into two beams $|\alpha\rangle$ $|\alpha\rangle$ \cite{hebing}, and then they are coupled to the photonic modes $a_{1}$ and $c_{1}$,
$a_{2}$ and $c_{2}$ through the XPM respectively. Correspondingly, the phase shifts induced by the couplings are $\theta$ and $2\theta$ in both beams.
Then we separate each path into two in terms of frequency by optical demultiplexers (OD) \cite{WDM}, all the upper paths correspond to frequency $\omega_{1}$ and can induce
a phase shift of $\theta$ on the upper coherent state, while all the under paths correspond to frequency $\omega_{2}$ and bring the same
phase shift to the under coherent state. Through an X homodyne measurement \cite{homodyne}, if the two coherent states have the same phase shift,
namely corresponding to the last two terms in both spatial mode and frequency DOFs, the state of the four-photon system becomes
\begin{eqnarray}
|\phi\rangle
&=&\frac{1}{2}(\alpha^2|HHHV\rangle-\beta^2|VVVH\rangle\nonumber\\
&&-\alpha\beta|HHVH\rangle+\alpha\beta|VVHV\rangle) \nonumber\\
&&\otimes(|a_{1}b_{1}c_{2}d_{2}\rangle+|a_{2}b_{2}c_{1}d_{1}\rangle)\nonumber\\
&&\otimes(|\omega_{1}\omega_{2}\omega_{2}\omega_{1}\rangle+|\omega_{2}
\omega_{1}\omega_{1}\omega_{2}\rangle)\; .\label{shi2}
\end{eqnarray}

\begin{figure}[h]
\centering
\includegraphics[width=8cm]{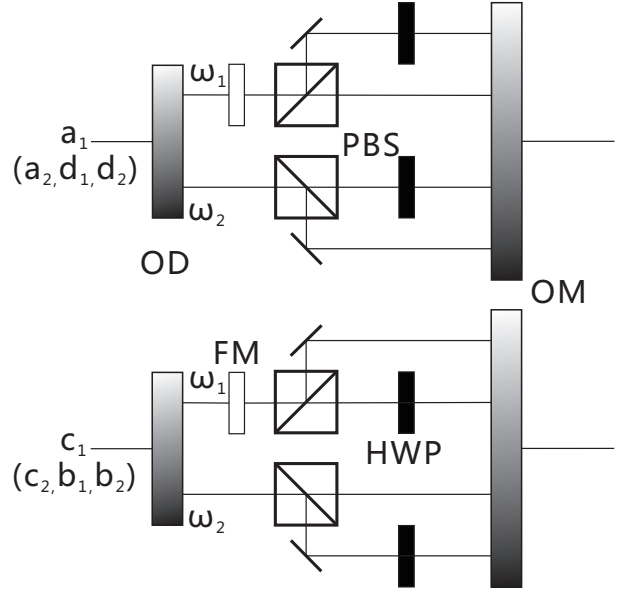}
\caption{Scheme of realizing entanglement transformation between polarization and frequency DOFs, and erasing frequency entanglement information of photon pair CD.
HWP represents a half-wave plate which is used to perform a bit-flip operation in polarization.} \label{fig2}
\end{figure}

Then we perform entanglement transformation between polarization and frequency DOFs by dint of the apparatuses shown in Figure 2.
After dividing each path into two with different frequency by OD, we use polarizing beam splitters (PBSs) and half-wave plates (HWPs)
to make the polarizations of photons A and D change into horizontal polarizations if their frequencies are $\omega_{1}$,
whereas if their frequencies are $\omega_{2}$, they will be vertical polarizations. In contrast, the transformations in the
polarizations of photons C and B are completely opposite to photons A and D. Furthermore, put an frequency multiplier (FM) in each path corresponding to $\omega_{1}$,
in order to turn the frequencies of the four photons all into the same, i.e. $\omega_{2}$. Whereupon the state of total system can be rewritten as:
\begin{eqnarray}
|\phi\rangle
&=&\frac{1}{2}(|HHHH\rangle+|VVVV\rangle)\nonumber\\
&&\otimes(|a_{1}b_{1}c_{2}d_{2}\rangle+|a_{2}b_{2}c_{1}d_{1}\rangle)\nonumber\\
&&\otimes|\omega_{2}\omega_{2}\omega_{2}\omega_{2}\rangle\; .\label{shi2}
\end{eqnarray}

Finally, we need only to extract the hyperentanglement of AB out of the four-body hyperentanglement.
After going through the devices shown in Figure 3, the corresponding state of CD in polarization and spatial mode becomes $|HH\rangle\otimes|c_{1}d_{1}\rangle$.
In this way, we obtain the maximally hyperentangled state of photon pair AB, $|\varphi\rangle_{AB}=\frac{1}{2}(|H\rangle|H\rangle+|V\rangle|V\rangle)
\otimes(|a_{1}\rangle|b_{1}\rangle+|a_{2}\rangle|b_{2}\rangle)$.
\begin{figure}[h]
\centering
\includegraphics[width=8cm]{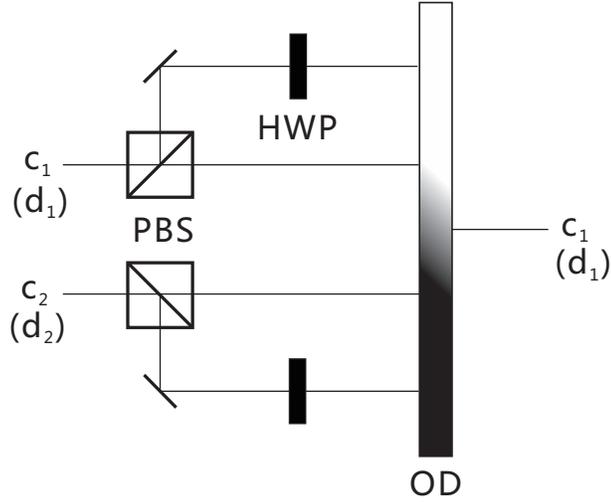}
\caption{Scheme of erasing entanglement informations of photon pair CD in polarization and spatial mode.} \label{fig3}
\end{figure}

Similarly, the other 15 cases also have the same result. In principle, as long as the two coherent states at Alice's side have the same phase shift,
this protocol succeeds with the probability $P=4|\gamma\delta\varepsilon\eta|^2$. Otherwise, it fails. Therefore, in practice, we only need Alice
to judge whether it succeeds or fails, and do not rely on postselection from both sides.

\section{Hyper-ECP with additional maximal frequency entanglement}
In this section, we assume that the polarization states and spatial mode states of the two-photon systems are both turned into
the following mixed forms after transmission:
\begin{eqnarray}
\rho_{p}&=&F_{1}|\psi_{p_{1}}\rangle\langle\psi_{p_{1}}|+F_{2}|
\psi_{p_{2}}\rangle\langle\psi_{p_{2}}|\nonumber\\
&&+F_{3}|\psi_{p_{3}}\rangle\langle\psi_{p_{3}}|+F_{4}|\psi_{p_{4}}
\rangle\langle\psi_{p_{4}}| \\
\rho_{s}&=&G_{1}|\psi_{s_{1}}\rangle\langle\psi_{s_{1}}|+G_{2}|
\psi_{s_{2}}\rangle\langle\psi_{s_{2}}|\nonumber\\
&&+G_{3}|\psi_{s_{3}}\rangle\langle\psi_{s_{3}}|+G_{4}|\psi_{s_{4}}
\rangle\langle\psi_{s_{4}}|\; ,\label{shi2}
\end{eqnarray}
where $F_{1}+F_{2}+F_{3}+F_{4}=1$, $G_{1}+G_{2}+G_{3}+G_{4}=1$ and
\begin{eqnarray}
|\psi_{p_{1}}\rangle=\alpha|HH\rangle+\beta|VV\rangle,\nonumber\\
|\psi_{p_{2}}\rangle=\alpha|HV\rangle+\beta|VH\rangle,\nonumber\\
|\psi_{p_{3}}\rangle=\alpha|HH\rangle-\beta|VV\rangle,\nonumber\\
|\psi_{p_{4}}\rangle=\alpha|HV\rangle-\beta|VH\rangle;\label{shi2}
\end{eqnarray}
\begin{eqnarray}
|\psi_{s_{1}}\rangle=\gamma|a_{1}b_{1}\rangle+\delta|a_{2}b_{2}\rangle,\nonumber\\
|\psi_{s_{2}}\rangle=\gamma|a_{1}b_{2}\rangle+\delta|a_{2}b_{1}\rangle,\nonumber\\
|\psi_{s_{3}}\rangle=\gamma|a_{1}b_{1}\rangle-\delta|a_{2}b_{2}\rangle,\nonumber\\
|\psi_{s_{4}}\rangle=\gamma|a_{1}b_{2}\rangle-\delta|a_{2}b_{1}\rangle.\label{shi2}
\end{eqnarray}
While the frequency state sill remains maximally entangled, i.e.
\begin{eqnarray}
|\psi_{f}\rangle=\frac{1}{\sqrt{2}}(|\omega_{1}\omega_{2}\rangle+|\omega_{2}\omega_{1}\rangle)\; .\label{shi2}
\end{eqnarray}

We also use two pairs of photons AB and CD in the new mixed states. Similar to the protocol introduced in preceding section,
the photons A and C belong to Alice,
and the other two B and D belong to Bob. The four parameters
$\alpha$, $\beta$, $\gamma$ and $\delta$ are unknown to Alice and Bob, and satisfy the relation
$|\alpha|^2+|\beta|^2=|\gamma|^2+|\delta|^2=1$.

\begin{figure}[h]
\centering
\includegraphics[width=8cm]{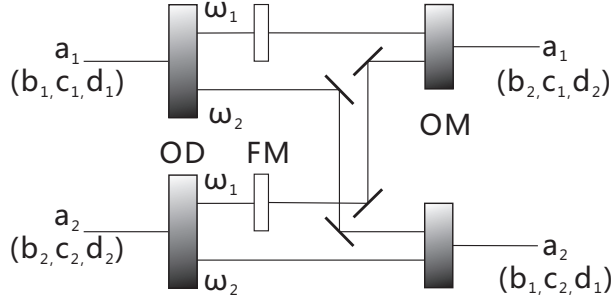}
\caption{Scheme of realizing entanglement transformation between spatial mode and frequency DOFs, and erasing frequency entanglement information of two photon pairs AB and CD.} \label{fig4}
\end{figure}

In this case, there will be 256 kinds of combinations. We take the following generic form as an example:
\begin{eqnarray}
|\phi\rangle&=&|\psi_{p_{1}}\rangle_{AB}|\psi_{s_{1}}
\rangle_{AB}|\psi_{f}\rangle_{AB}\nonumber\\
&&\otimes|\psi_{p_{2}}\rangle_{CD}|\psi_{s_{2}}\rangle_{CD}|\psi_{f}\rangle_{CD} \nonumber\\
&=&(\alpha|HH\rangle+\beta|VV\rangle)(\alpha|HV\rangle+\beta|VH\rangle) \nonumber\\
&&\otimes(\gamma|a_{1}b_{1}\rangle+\delta|a_{2}b_{2}\rangle)\nonumber\\
&&(\gamma|c_{1}d_{2}\rangle+\delta|c_{2}d_{1}\rangle) \nonumber\\
&&\otimes\frac{1}{2}(|\omega_{1}\omega_{2}\rangle+|\omega_{2}\omega_{1}\rangle)\nonumber\\
&&(|\omega_{1}\omega_{2}\rangle+|\omega_{2}\omega_{1}\rangle)\; .\label{shi2}
\end{eqnarray}

To begin with, we can change the spatial mode states of the two pairs into maximal entanglement respectively with the quantum circuit shown in Figure 4.
Alice and Bob let the two paths of each photon link with an OD and be respectively separated into different paths again according to frequency.
As shown, the paths with the same frequency will be merged into the same path by an optical multiplexer (OM). Here we also need to
erase the information of frequency entanglement by frequency multipliers. Hence, the state of the four photons becomes
\begin{eqnarray}
|\phi\rangle&=&\frac{1}{2}(\alpha^2|HHHV\rangle+\beta^2|VVVH\rangle\nonumber\\
&&+\alpha\beta|HHVH\rangle+\alpha\beta|VVHV\rangle) \nonumber\\
&\otimes&(|a_{1}b_{1}\rangle+|a_{2}b_{2}\rangle)(|c_{1}d_{1}
\rangle+|c_{2}d_{2}\rangle)\nonumber\\
&&\otimes|\omega_{2}\omega_{2}\omega_{2}\omega_{2}\rangle\; .\label{shi2}
\end{eqnarray}

\begin{figure}[h]
\centering
\includegraphics[width=8cm]{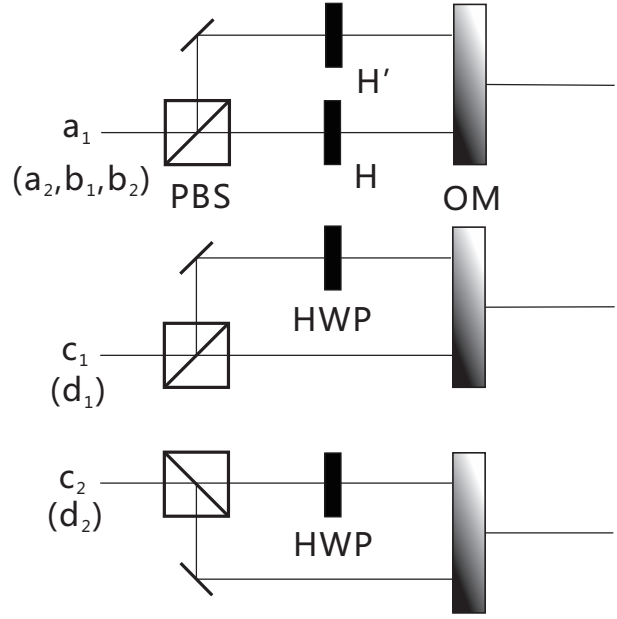}
\caption{Scheme of transforming the polarization of photons A and B into superposition state, and
forming an EPR-like entangled state between polarization and spatial mode DOFs of photon pair CD.
H (H') represents the operation $|H\rangle (|V\rangle)\rightarrow\frac{|H\rangle+|V\rangle}{\sqrt{2}}$.} \label{fig5}
\end{figure}

Next, as shown in Figure 5, both Alice and Bob perform H and H' operations on the horizontal and vertical polarization of photons A and B respectively.
As a result, horizontal and vertical polarization are both transformed into superposition state $\frac{|H\rangle+|V\rangle}{\sqrt{2}}$.
And for the photon C (D), we transform its polarization into $H$ and $V$, according to its spatial modes $c_{1}$ ($d_{1}$) and $c_{2}$ ($d_{2}$) respectively
with PBSs, HWPs and OMs. Then the state of the whole system can now be written as
\begin{eqnarray}
|\phi\rangle&=&\frac{1}{4}(|H\rangle+|V\rangle)(|H\rangle+|V\rangle)\nonumber\\
&&\otimes(|a_{1}b_{1}\rangle+|a_{2}b_{2}\rangle)\nonumber\\
&&\otimes(|HHc_{1}d_{1}\rangle+|VVc_{2}d_{2}\rangle)\nonumber\\
&&\otimes|\omega_{2}\omega_{2}\omega_{2}\omega_{2}\rangle\; .\label{shi2}
\end{eqnarray}

Subsequently, as shown in Figure 6, both Alice and Bob exploit another quantum nondemolition detector (QND2) composed of PBSs and cross-Kerr nonlinear medium.
The cross-Kerr nonlinearity will make the upper coherent beam $|\alpha\rangle$ pick up a phase shift $\theta$, if the polarization of photon A (B) is $H$ ($H$)
or photon C (D) is in the mode $c_{2}$ ($d_{2}$). While if the polarization of photon A (B) is $V$ ($V$) or photon C (D) is in the mode $c_{1}$ ($d_{1}$),
there will be a a phase shift $\theta$ in the under coherent beam $|\alpha\rangle$. After performing homodyne measurements on both sides and transforming the state of CD
into $|HH\rangle\otimes|c_{1}d_{1}\rangle$, if the differences between
phase shift of upper $|\alpha\rangle$ and phase shift of under $|\alpha\rangle$ from Alice and Bob are both 0 or $2\theta$ (homodyne detection can't distinguish plus and minus \cite{plus,minus}),
the selected terms constitute the maximally hyperentangled state of photon pair AB, $|\varphi\rangle_{AB}=\frac{1}{2}(|H\rangle|H\rangle+|V\rangle|V\rangle)
\otimes(|a_{1}\rangle|b_{1}\rangle+|a_{2}\rangle|b_{2}\rangle)$. If one of the differences is 0, the other is $2\theta$, the state of AB becomes
$|\varphi\rangle_{AB}=\frac{1}{2}(|H\rangle|V\rangle+|V\rangle|H\rangle)
\otimes(|a_{1}\rangle|b_{1}\rangle+|a_{2}\rangle|b_{2}\rangle)$. And then we can get the ideal
maximally hyperentangled state by performing the bit-flipping operation $\sigma_{x}=|H\rangle\langle V|+|V\rangle\langle H|$ on the photon A or B.

\begin{figure}[h]
\centering
\includegraphics[width=8cm]{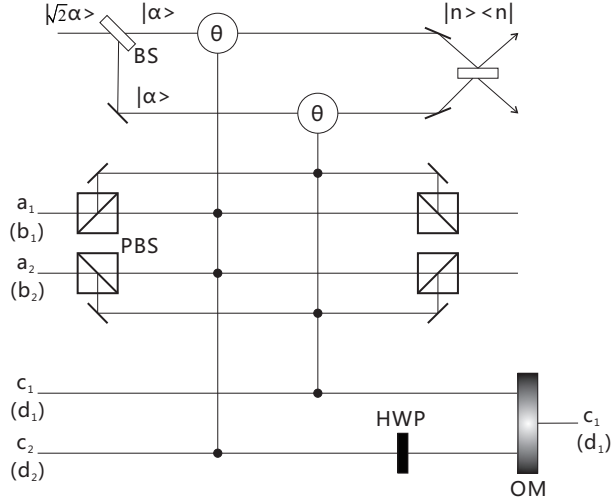}
\caption{$QND_{2}$. If the results of homodyne measurements from Alice and Bob are same, the state of photon pair AB is projected into the maximally hyperentangled state
$|\varphi\rangle_{AB}=\frac{1}{2}(|H\rangle|H\rangle+|V\rangle|V\rangle)
\otimes(|a_{1}\rangle|b_{1}\rangle+|a_{2}\rangle|b_{2}\rangle)$. If the results are different, only a bit-flip operation
$\sigma_{x}=|H\rangle\langle V|+|V\rangle\langle H|$ on the photon A or B is needed to obtain the state $|\varphi\rangle_{AB}$.} \label{fig6}
\end{figure}

For other combinations, in the same way, we can purify arbitrary mixed states in polarization and spatial mode,
and distill target systems in the maximally hyperentangled state from whole systems in the patially hyperentangled state.
Alice and Bob can determinately obtain the expected nonlocal states just through local operation and classical communication
(LOCC), i.e. success probability is $100\%$.

\section{Conclusions}

In summary, we have proposed two hyper-ECPs for two-photon systems in partially hyperentangled unknown state,
resorting to linear optical instruments and cross-Kerr nonlinearities. In the first situation,
the entangled states in all the three DOFs turn to be in the form of partial entanglement, while only the polarization part may suffer from bit-flip and phase-flip errors during transmission.
The two nonlocal parties use two arbitrary photon pairs in the unknown hyperentangled mixed state to correct errors in polarization and extract maximal entanglement in polarization
and spatial mode. It is obvious that whether this ECP may
succeed or not only depends on the result of QND1 (see Figure 1)
placed on the side of Alice. The success probability changes with
the parameters of the partially entangled states in spatial mode and frequency. Besides, it is easy to see when the parameters of the
partially polarization entangled states of the two pairs are different, even for non-entangled states, the ECP is also applicable. In the second situation, both of the entangled states
in polarization and spatial mode are likely to suffer from bit-flip and phase-flip errors, and become less-entangled, but the frequency entanglement remains intact. The two parties can get a pair of photons with maximal entanglement in both polarization and spatial
mode DOFs by LOCC and the consumption of another photon pair and
their frequency entanglements determinately.
Especially, for the two photon pairs, all the four parameters of the partially hyperentangled states can be arbitrary (including 0) and different from each other.
Theoretically, our ECPs can be simply expanded to the situation with more photons or other particles, thus they are generally applicable.

In addition, both of the two ECPs are based on the cross-Kerr nonlinearity, but traditional cross-Kerr nonlinearity materials offer only weak coupling at the single-photon level and no overall phase shift can be induced by cross-phase modulation \cite{Skerr1,Skerr2}. However, electromagnetically induced transparency has been demonstrated that it may afford a strong cross-Kerr effect between weak optical fields \cite{EIT1,EIT2}. Moreover, Munro {\it et al.} \cite{Munro} showed a new
method of the realization of quantum non-demolition measurements by virtue of successive weak cross-Kerr interactions. In principle,
our non-demolition measurements here only involve the parity check for two photons, and the strong cross-Kerr nonlinearity is not required.

In a practical application, our ECPs can be realized with current technology and will greatly improve the efficiency and fidelity of long-distance quantum communication, which enable us to take full
advantage of the superiority of hyperentanglement in the future.

%%%%%%%%%%%%%%%%%%%%%%%%%%%%%%%%%%%%%%%%%%%%%%%%%%%%%%%%%%%%%%%%%%%%%%%%%%%
\vspace{.3cm} {\bf Acknowledgments}

This work was supported in part by the National Natural Science
Foundation of China(NSFC) under the grants 11121092, 11175249 and 11375200.
%%%%%%%%%%%%%%%%%%%%%%%%%%%%%%%%%%%%%%%%%%%%%%%%%%%%%%%%%%%%%%%%%%%%%%%%%%%

\end{document}